\documentclass[12pt,epsfig,rotate]{article}
\usepackage{epsfig}
\usepackage{wrapfig}
\usepackage{graphicx}
\begin{document}

\setcounter{page}{1}

\begin{center}
{\bf
THE STRUCTURE EFFECTS IN POLARIZATION AND CROSS SECTION
IN INELASTIC A({\it p, p'})X REACTION WITH THE $^{40}$Ca
AND $^{12}$C NUCLEI AT 1~GeV}\\

\vspace*{9mm}
{\bf O.V.~Miklukho, A.Yu.~Kisselev, G.M.~Amalsky, V.A.~Andreev,
G.E.~Gavrilov, A.A.~Izotov, N.G.~Kozlenko, P.V.~Kravchenko, M.P.~Levchenko,
D.V.~Novinskiy, A.N.~Prokofiev, A.V.~Shvedchikov, S.I.~Trush,
A.A.~Zhdanov}\\
\vspace*{5mm}
{\it B.P.~Konstantinov Petersburg Nuclear Physics Institute, National Research Centre
Kurchatov Institute, 
Gatchina, 188300 Russia}\\
\end{center}

\vspace*{10mm}
The polarization of the secondary protons  in the
inelastic  ({\it p, p'}) reaction on the $^{40}$Ca and $^{12}$C nuclei at the initial
proton energy 1 GeV was measured in a wide range of the scattered
proton momenta  at a laboratory angle $\Theta$=21$^\circ$.
The cross sections of the reaction were measured as well. The outgoing protons
from the reaction were detected using a magnetic spectrometer equipped with
a multiwire-proportional chambers polarimeter. A structure in the polarization
and cross section data, related probably to scattering off the nucleon correlations
in the nuclei, was observed.
\\
\\
{\bf Comments:} 13~pages, 5~figures, 7~tables. \\
\\
\\
\\
{\bf Category:} Nuclear Experiment (nucl--ex)\\

\newpage
\section{Introduction}
 ---This  work is a part of the experimental program in the framework of which
the effects of nucleon clusterization in nuclear matter is
studied at the PNPI synchrocyclotron with the 1~GeV proton beam \cite{Miklukho2011, Miklukho2015}.
Earlier, in the first inclusive experiment, the scattered proton polarization 
in the reaction $^{40}$Ca({\it p,~p'})X at $\Theta$=21$^\circ$ was measured \cite{Miklukho2011}.
At a proper secondary proton momentum, when scattering off the $^{4}$He-like
nucleon cluster (nucleon correlation (NC)) in the  $^{40}$Ca nucleus  could dominate,
the measured polarization was found to be  close to that
in the free elastic proton-$^{4}$He scattering. We investigated in details 
the polarization in the reaction and observed a structure
in the experimental data \cite{Miklukho2015}. The latter could be related to scattering off
the multi-nucleon correlations in the nucleus \cite{Blokhintsev1957, CLAS2006}.   

In this paper we present the results of the experiment, in which two inclusive
reactions $^{40}$Ca({\it p, p'})X and $^{12}$Ca({\it p, p'})X 
at the scattering angle of the secondary protons $\Theta$=21$^\circ$ were investigated.
In this experiment, besides the polarization of the final protons,
we also measured the differential cross section of the reactions. 
These measurements were performed
in a wide range of the scattered proton momentum $K$ ($K$~=~1370$\div$1670~MeV/c).
Note that the momentum corresponding to a maximum of the quasielastic $pN$ peak is
close to 1480~MeV/c. 
The data were obtained 
in narrow momentum intervals ($\simeq$~10~MeV/c) and with a small gap between the
intervals ($\simeq$~10~MeV/c). 
Of a special interest was to make the measurements at the $K$~$>$~1530~MeV/c up to the momentum
corresponding to the excited level of the nucleus under investigation.
In this region, since the NC are more
massive than nucleons, the quasi-elastic ({\it p, p'}~NC) reactions
(the elastic scattering off the NC in nuclear medium) are kinematically preferable.
At $K$~$>$~1580 MeV/c the scattering off the independent (uncorrelated) 
nuclear  nucleons is strongly suppressed,
since they have a minimal momentum larger than the Fermi momentum $K_F$ ($\approx$~250~MeV/c) [2].
\clearpage
 The general layout of the experimental setup  is presented in Fig.~1.

\begin{figure}
\centering\epsfig{file=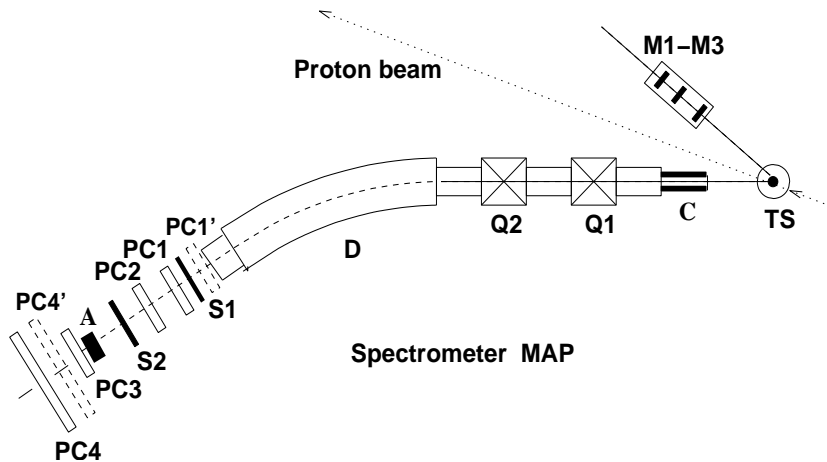,width=.75\textwidth=100mm,height=60mm}
\caption{\small The experimental setup. TS is the target of the MAP spectrometer;
Q1$\div$Q2 are the magnetic quadrupoles; D is the dipole
magnet; C1 is the collimator; S1$\div$S2 and M1$\div$M3 are
the scintillation counters; PC1$\div$PC4, PC1', PC4'
and A are the proportional
chambers and the carbon analyzer of the MAP
polarimeter, respectively.}
\end{figure}

\section{Experimental method}
The proton beam of the PNPI synchrocyclotron was focused
onto the  target TS of the magnetic  spectrometer MAP. The beam intensity 
was monitored by the scintillation telescope M1, M2, M3. 
The diameter of the beam spot 
on the target was $\simeq$~25~mm.
Large CH$_2$ and C targets, CH$_2$ foils for the setup
calibration, and  small $^{12}$C and $^{40}$Ca targets for the main measurements   
were used in the experiment (Table~1). 
\\
\begin{table}
\caption{Target parameters}
\label{table:targets}
\begin{center}
\begin{tabular}{l|c|c|c}
\hline
Target        &  Dimensions, mm     & Isotope concentration, $\%$ & Density, g/cm$^3$ \\
\hline
              & thickness x width x height &                 & \\
\hline
CH$_2$              & 4 x 15 x 70        &                   & 1.0 \\
C                   & 4 x 15 x 70        & 98.9              & 1.6 \\
CH$_2$ foil         & 0.1 x 4 x 10       &                   & 1.0 \\
\hline
$^{12}$C            & 4 x 7 x 10         & 98.9              & 1.6  \\          
$^{40}$Ca           &  4 x 7 x 10        & 97.0              & 1.55 \\         
\hline
\end{tabular}
\end{center}
\end{table}                                   
\\
\\
The spectrometer was used to measure the momenta of the secondary protons  
from the inclusive  ({\it p,~p'}) reaction as well as their polarization.
The momentum of the proton was determined using the coordinate information 
from the proportional chambers PC1-X and PC2-X.
The momentum resolution of the spectrometer in this experiment
was $\pm$ 2.5 MeV/c. This value was estimated by measuring the width of
the clearly separated 2$^+$ excited level in the ({\it p, p'}) reaction 
with the $^{12}$C nucleus at the scattering angle 21$^\circ$ under investigation (Fig.~2).
In Fig.~2 we also observed a peak which can be identified as the 1$^+$ excited level predicted 
in \cite{Viollier1975}.

\begin{figure}
\centering\epsfig{file=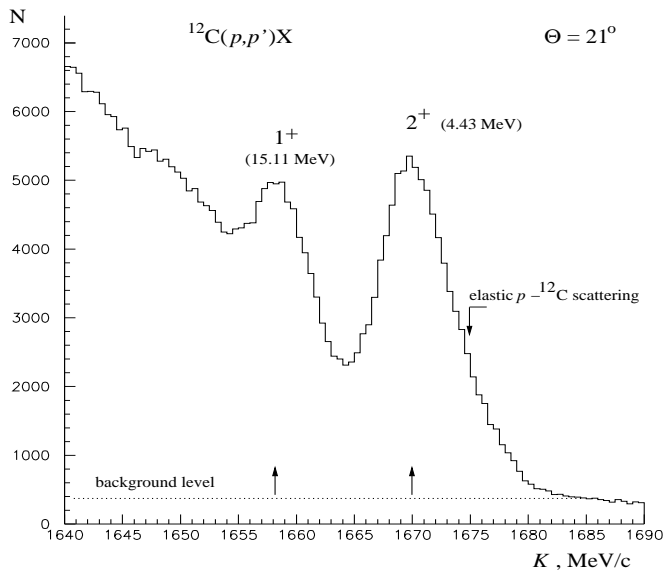,width=.6\textwidth,height=75mm}
\caption{\small Momentum distribution in the inclusive reaction $^{12}$C($p, p'$)X  at a 
       scattering angle $\Theta$=21$^\circ$. }
\end{figure}
The polarization of the final protons was found from an azimuthal asymmetry of the proton scattering
off the carbon analyzer A, using the track information from the proportional chambers 
(PC1$\div$P4 and PC1', PC4') of the polarimeter \cite{Miklukho2013}.
The average analyzing power of the polarimeter was calculated using the parametrization 
$A(K,\theta_s)$  from \cite{Fedorov1979}. 
 
The main parameters of the MAP spectrometer and the
polarimeter are listed in Tables~2 and~3, respectively.\\
\begin{table}
\caption{Parameters of the magnetic spectrometer}
\label{table:spectrometer}
\begin{center}
\begin{tabular}{l|c}
\hline
Maximum particle momentum, [GeV/c]                     & 1.7\\
Horizontal angle acceptance $\Delta \Theta_H$, [deg]   & 0.8\\
Vertical angle acceptance $\Delta \Theta_V$, [deg]     & 1.9\\
Solid angle acceptance $\Omega$, [sr]                  & $4.0\cdot 10^{-4}$\\
Momentum acceptance $\Delta K/K$, [$\%$]               & 8.0\\
Dispersion in the focal plane $D_f$, [mm/$\%$]         & 22.0\\
Momentum resolution (FWHM), [MeV/c]                    & $\sim 5.5$\\
\hline
\end{tabular}
\end{center}
\end{table}
\\
\begin{table}
\caption{Polarimeter parameters}
\label{table:polarimeter}
\begin{center}
\begin{tabular}{l|c}
\hline
Carbon block thickness, [mm]     & 155\\
Polar angular range, [deg]       & 3$\div$16\\
Average analyzing power          & $\geq$ 0.2\\
Efficiency, [$\%$]               & $\sim$ 5\\
\hline
\end{tabular}
\end{center}
\end{table}
The calibration of the analyzing power of the
polarimeter was carried out using the {\it pp} elastic scattering
polarization data obtained in this experiment. For the  calibration in a wide range of the
 secondary proton energy
we performed the polarization measurements with the polyethylene 
and carbon targets (the large CH$_2$ and C targets, Table~1) at different angular 
($\Theta$~=~13.5$^\circ$ $\div$ 23 $^\circ$) 
and proper momentum settings of the spectrometer. The observed values 
of the {\it pp}~polarization were 
compared with the predictions in the framework of  the phase-shift analysis \cite{Lehar1978}
and a correction for the analyzing power of the polarimeter has been done. 
The uncertainty of the calibration was included in the total error
of the polarization measurement. 

The relative differential cross section of the reactions 
$\sigma^{incl}$~=~$\frac{d^2\sigma}{d\Omega dK}$
was found from the momentum spectra obtained at different momentum settings of the
spectrometer. The monitor number and efficiency of the proportional chamber PC2-X
for each momentum setting were taken into account. To subtract a background,
the measurements with an empty target and with the 100~$\mu$m tungsten  string 
supports of a target were done. An absolute normalization of the cross section 
in the reaction with the $^{12}$C nucleus was made with the large CH$_2$ target (Table~1) 
in a momentum range near the quasi-elastic scattering peak maximum. 
The cross section in the {\it pp} elastic scattering
was calculated in the framework of the phase-shift analysis \cite{Lehar1978}.
At the scattering angle $\Theta$~=~21$^\circ$ a value of 2.92~mb/sr (in the centre-of-mass system) 
was obtained.
To make this normalization for the reaction with the $^{40}$Ca nucleus, we
made the measurements with the $^{40}$Ca target, with the $^{40}$Ca target + 
additional target CH$_2$-foil (Table~1), and again with the $^{40}$Ca target.
These data allowed us to estimate the admixture of hydrogen in the superficial
layer of the $^{40}$Ca target.
The relative systematic errors $\delta\sigma^{incl}$/$\sigma^{incl}$  
of the cross section ($\sigma^{incl}$) normalization for the reactions
$^{12}$C({\it p,~p'})X and $^{40}$Ca({\it p,~p'})X   
were $\pm$~1.5 $\%$ and $\pm$~3.5, 
respectively. An uncertainty of the
cross section calculations in the elastic $pp$ scattering \cite{Lehar1978} was not included.

\section{Experimental results and discussion}

\begin{figure}
\centering\epsfig{file=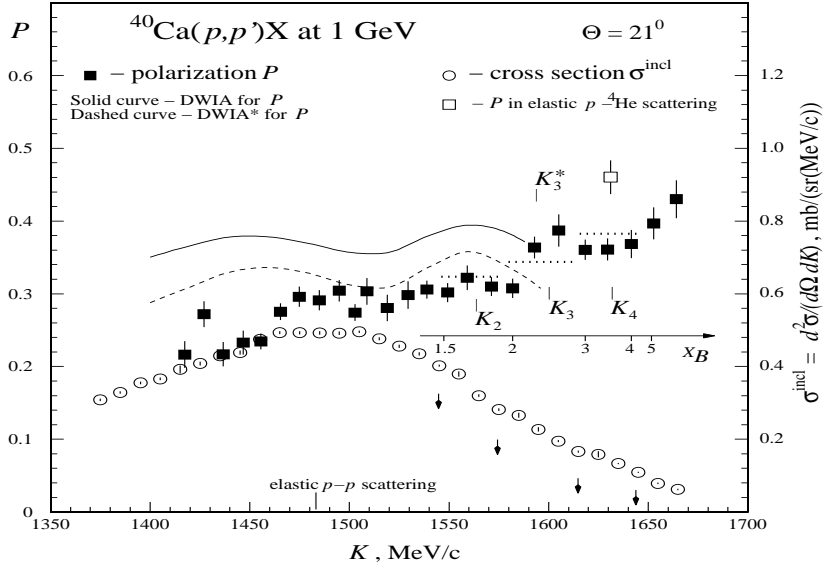,width=.75\textwidth,height=75mm}
\caption{\small Polarization $P$ of the protons scattered  at an angle $\Theta$~=~21$^\circ$ 
(black squares) in the inclusive reaction $^{40}$Ca($p, p'$)X versus the
secondary proton momentum $K$. The circles correspond to the 
differential cross sections $\frac{d^2\sigma}{d\Omega dK}$ measured
in the reaction. Solid and dashed curves are a result of the 
polarization calculations in the framework of the DWIA and DWIA*, respectively.
The empty square corresponds to the polarization
in the elastic $p-^4$He  scattering \cite{Miklukho2006}. The dotted lines cover 
the $K$ intervals II, III, IV  and the Bjorken variable scale are defined in 
the text.}
\end{figure}
\begin{figure}
\centering\epsfig{file=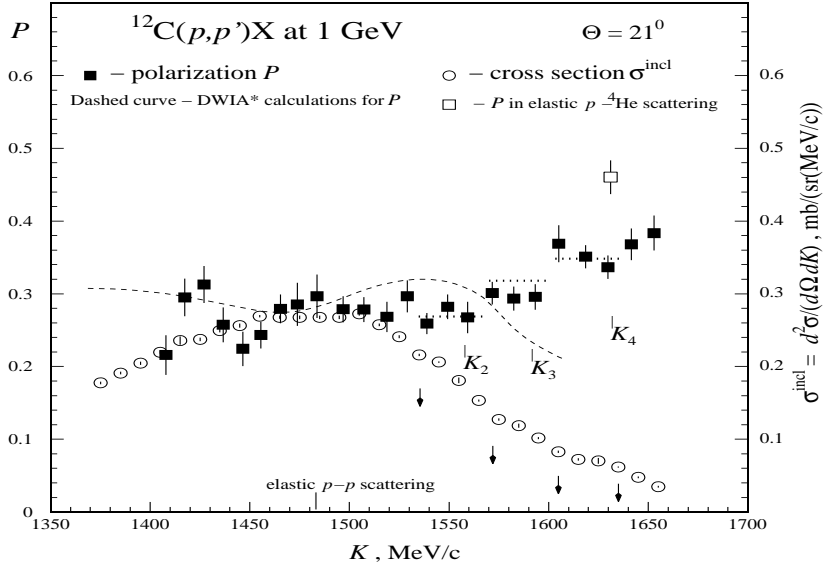,width=.75\textwidth,height=75mm}
\caption{\small  Polarization $P$ of the protons scattered  at an angle $\Theta$~=~21$^\circ$
(black squares) in the inclusive reaction $^{12}$Ca($p, p'$)X versus the
secondary proton momentum $K$. The circles correspond to the
differential cross sections $\frac{d^2\sigma}{d\Omega dK}$ measured
in the reaction. Dashed curve is a result of the polarization calculations
in the framework of the DWIA*.
The empty square corresponds to the polarization
in the elastic $p-^4$He scattering \cite{Miklukho2006}. The dotted lines cover
the $K$ intervals II, III, and IV, defined in the text.}
\end{figure}
In Fig.~3 and Fig.~4 the measured polarizations $P$ (black squares) and 
cross sections $\frac{d^2\sigma}{d\Omega dK}$ (circles)
in the reactions $^{40}$Ca({\it p,~p'})X and $^{12}$C({\it p,~p'})X 
are plotted versus the scattered proton momentum $K$.
Small errors of the cross section measurements
are presented inside the circles.                     
The experimental data are also given in Tables~4$\div$7. 
The empty square corresponds to an estimate of the polarization 
in the elastic $p-^4$He  scattering \cite{Miklukho2006}.
The solid curve in Fig.~3 presents the polarization calculated
in the framework of a spin-dependent Distorted Wave Impulse Approximation
(DWIA) \cite{Andreev2004}. The dashed curves in Fig.~3 and Fig.~4 are the result 
of the calculations in the framework of the DWIA taking into account 
the relativistic distortion of the nucleon spinor in nuclear medium (DWIA*) 
\cite{Andreev2004, Horowitz1986}. In this approach 
the proton scattering off the independent nuclear nucleons was taken into account only.
The calculations were performed using the THREEDEE code \cite{Chant1983}. 

As seen from Figs.~3,~4 in the region of the $K$~$>$~1530~MeV/c a drop of 
the cross section slows down at the momenta 
close to these marked by arrows. The momentum
intervals between the adjacent arrows are indicated in the figures as the dotted line
segments in an area of the polarization data. Let us denote these momentum ranges
as II, III, and IV in the direction of momentum growth.
Note that the onset of each momentum
interval (II, III, and IV)  for the $^{40}$Ca nucleus
is shifted with respect to that for the $^{12}$C nucleus by $\sim$~ 5$\div$10~MeV/c
towards higher values.     
The value of the polarization in
the momentum ranges is  practically constant excluding 
the interval III for the $^{40}$Ca data.  
The polarization increases from the interval II to interval IV. At momenta $K~>~$1580~MeV/c
the large values of the polarization and cross section \cite{Miklukho2015}
can not be explained only by the proton scattering 
off the uncorrelated nuclear nucleons. Possible, such behaviour of  polarization and cross section 
in the momentum ranges II, III, and IV can be related to a proton quasielastic
scattering off the two-nucleon, three-nucleon, and four-nucleon correlations.
The value of the polarization in the proton interaction
with a NC can depend on the number and isospin properties of nucleons in the correlation.
According to \cite{Faisner1959}, larger values of the secondary proton polarization
can be observed in the elastic scattering off a light nucleus 
in comparison with that in the scattering off independent nuclear nucleons.

The calculated final proton momenta $K_2$,~~$K_3$~($K^*_3$),~~and~~ $K_4$ corresponding 
to the maxima 
of the quasi-elastic peaks in the $^{40}$Ca({\it p,~p'}~NC)X and $^{12}$C({\it p,~p'}~NC)X 
reactions on the stationary NC consisting of two, three, and four nucleons 
are shown in Figs.~3,~4. In these calculations the masses of real light nuclei with 
simple structure 
$^2$H, $^3$He ($^3$H), and $^4$He were used as the NC masses.
The residual nuclei (X) in the reactions were assumed to be in a ground state.
As seen in Fig.~3 the momenta $K_2$ (1563~MeV/c), $K_3$ (1599~MeV/c), $K^*_3$ (1593~MeV/c), 
and $K_4$ (1631~MeV/c), and in Fig.~4 the momenta $K_2$ (1557~MeV/c), 
$K_3\approx~K^*_3$ (1591~MeV/c), and $K_4$ (1631~MeV/c)
are within the momentum intervals II, III, and IV, respectively. 
This observation stays true
if the NC masses were smaller (due to
the nuclear medium modification \cite{Horowitz1986}) than the mass of
the corresponding free light nucleus. A $\sim$~10~$\%$ decrease of the NC masses 
reduces  the values of the momenta $K_2$, $K_3$ ($K^*_3$), and  $K_4$
by $\sim$~12~MeV/c, $\sim$~8~MeV/c, and $\sim$~6~MeV/c, respectively.
Note here that
a high momentum range, just following  the momentum interval IV,
possibly corresponds to  quasi-elastic scattering off the residual nuclei X
of the reactions considered above. 
 
The DWIA* calculations show that the contribution 
from the quasi-elastic scattering
off the uncorrelated nucleons in the momentum interval II is rather large \cite{Miklukho2015}. 
At $K$~$>$~1580 MeV/c, including the momentum intervals 
III and IV, this contribution is essentially suppressed since the nuclear nucleons have momenta
 higher than the Fermi momentum $k_F~\approx$~250~MeV/c.
The polarizations measured 
in the momentum interval IV ($P_{IV}$) in the scattering 
off the $^{40}$Ca and  $^{12}$C nuclei (see Figs.~3,~4)
are practically the same 
($P_{IV}$(Ca)~=~0.363~$\pm$~0.009 and $P_{IV}$(C)~=~0.348~$\pm$~0.010). 
The polarization $P_{IV}$  is less than that ($P_{^4{He}}$) 
in the free elastic $p-^4$He scattering 
(empty square) \cite{Miklukho2006}. This can be related
to a modification of proton interaction with the four-nucleon cluster in nuclear medium 
\cite{Miklukho2013}.
The relative difference of these polarizations ($P_{^4{He}}$~-~$P_{IV}$)/$P_{^4{He}}$ 
~$\sim$~0.2  is close to that  of the 
polarizations ($P_{DWIA^*}$ and $P_{DWIA}$) calculated in the DWIA* and DWIA approximations 
(Fig.~3) for the quasi-elastic 
scattering off the uncorrelated nucleons ($P_{DWIA}$~-~$P_{DWIA^*}$)/$P_{DWIA}$ 
~$\sim$~0.15 at $K~\approx$~1580~MeV/c.

The widths ($\Delta K$) of the momentum intervals II $\div$ IV are not determined by the
horizontal angular acceptance of the spectrometer alone ($\Delta\Theta_h\sim$~1$^\circ$).
The main contribution to the $\Delta K$ can come from a motion of the NC in the nucleus.
For instance, if the scattering occur off 
the four-nucleon  correlations at rest in the nucleus then the $\Delta K$ width of
the momentum interval IV would be equal to $\sim$~5.5~MeV/c. This value is about
4.5 times less than that estimated from this experiment $\Delta K$ $\sim$~25~MeV/c. 
\begin{figure}
\centering\epsfig{file=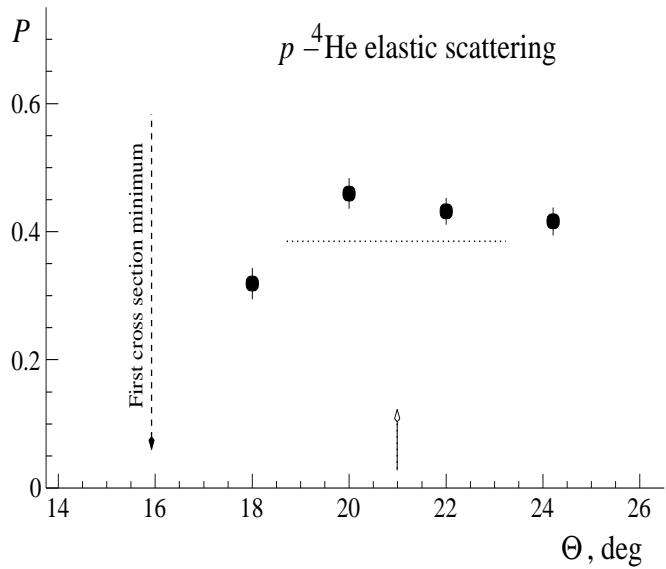,width=.6\textwidth,height=75mm}
\caption{\small Angular dependence of the polarization  
in the elastic $p-^4$He scattering at 1~GeV \cite{Miklukho2006}. 
The dotted line covers the effective angular acceptance
observed in the inclusive ($p, p'$) reaction.  }
\end{figure}
So, due to a motion of the NC,
the effective angular acceptance essentially 
increases ($\Delta\Theta_h\sim$~4.5$^\circ$). This enables us to observe
the polarization angular distribution in scattering from the NC within
the momentum interval IV. In Fig.~5 the polarization angular distribution 
measured in the free elastic $p-^4$He scattering is shown \cite{Miklukho2006}. The dotted line   
segment corresponds to the effective angular acceptance seen in the inclusive
reaction in the momentum interval IV. According to the data in the figure, we can expect that the 
polarization momentum dependence in the proton scattering
off the four-nucleon correlations in the nuclei can  also be close to uniform.
A growth of the polarization in the momentum interval III in the scattering off the $^{40}$Ca nucleus
(Fig.~3) is possibly related to that the momentum regions of scattering off
the $^3$He- and $^3$H-like correlations overlap only partially. This follows 
from a noticeable difference
of the momenta $K_3$ and $K^*_3$ mentioned above, being  $K_3$~$>$~$K^*_3$  
(for the $^{12}$C data these
momenta are almost the same). We suppose that the polarization in the elastic
scattering off the $^3$He nucleus is essentially larger than that off the $^3$H nucleus.
So, at momenta close to the end of the interval III, the polarization is possibly determined by
the scattering off the $^3$He-like correlations. At  $K~<~K^*_3$ the
processes of scattering off these three-nucleon correlations are mixed.
For reliable verification of the above supposition, the polarization calculations in the framework
of the Glauber's multiple nucleon-nucleon scattering theory \cite{Glauber1959} should be done.
We were only based on the fact that at the initial proton energy 1 GeV the polarization
in the elastic scattering off neutron is about 25 $\%$ less than that 
in the proton-proton scattering, and the number of neutrons in  $^3$He 
is smaller than in the $^3$H nucleus.

We would like to make some remarks about measured results 
in the momentum range 1420 MeV/c~$< K <$~1530 MeV/c 
(denote this range as I) covering the momentum ($K~\sim$~1480~MeV/c)
corresponding to a maximum of the $pN$ quasi-elastic peak (see Figs. 3, 4).
In the range I, where the cross section of the inclusive ($p, p'$) reaction
has large values and depends smoothly on the $K$, a contribution from the
multi-step processes of knocking out nucleons from a nucleus,
can be noticeable \cite{Wallace1985}. The outgoing proton momentum in 
these processes decreases as compared with that in the one-step
($p, p'$) reaction under investigation. Due to this effect a shape of
the quasi-elastic peak can be distorted. 
It is important to note here,
in the momentum
region $K >$~1530 MeV/c, where the cross section dips rapidly with a
growth of the $K$,
the multi-step processes are essentially unable to distort the 
momentum distribution measured in the reaction. 
The DWIA* predictions of the outgoing proton polarization 
for the reaction $^{12}$C($p, p'$)X (Fig. 4) are in a good consent with the
experimental data in a narrow region around the momentum $K~\sim$~1480~MeV/c.  
The variation of the measured polarization in the range I is apparently a combined
effect of the multi-step reactions and a discrete energy-shell
structure of the $^{12}$C  nucleus.

Last we make a comment on the 
kinematics of the present ({\it p, p'}) experiment.
In the momentum range 1480~$\div$~1650~MeV/c  the value of the transferred
four-momentum $Q$ stays almost constant and is equal to $\approx$ 600~MeV/c.
The latter value is about two times higher than that of the Fermi momentum. 
So, the Bjorken kinematical variable
$x_B = \frac{Q^2}{2m\nu}$  is only determined by the energy transfer
$\nu$ (where $m$ is nucleon mass). In Fig.~3, there is an additional
horizontal scale for  $x_B$ indicated. As seen from the figure, for the
reaction with the $^{40}$Ca nucleus  the momentum intervals II, III, and IV
correspond to the $x_B$ intervals 1.5~$< x_B <$~2, 2~$< x_B <$~3, and
3~$< x_B <$~4, respectively.
Due to the above mentioned difference in the momentum interval onsets
for the $^{12}$C and $^{40}$Ca nuclei
(it is possible related to a large mass difference of the nuclei),
the corresponding $x_B$ intervals in the reaction $^{12}$C($p, p'$)X 
are 1.4~$< x_B <$~1.8, 1.8~$< x_B <$~2.5, and 2.5~$< x_B <$~3.7.
It is interesting to note that in the JLAB unpolarized
($e,e'$) experiment at E$_e$ $\sim$ 4.6 GeV and $Q^2$~$>$~1.4~GeV$^2$/c$^2$,
the effects from the two-nucleon and three-nucleon correlations in the
cross section were observed in the $x_B$ ranges 1.5~$< x_B <$~2 and
2.25~$< x_B <$~2.8, respectively \cite{CLAS2006}.

\section{Summary}

The polarization of the secondary protons  in the
inelastic  ({\it p, p'}) reaction with the $^{40}$Ca and $^{12}$C nuclei 
and the cross section of these reactions were investigated 
at the 1~GeV initial proton energy and the scattering angle $\Theta$~=~21$^\circ$.
The data were obtained in a wide range of the scattered proton momentum $K$
covering the $pN$ quasi-elastic peak and a high momentum region ($K >$ 1530~MeV/c) up to
the momentum corresponding to the exited levels 
of the nucleus under investigation. The measurements were done
in narrow momentum intervals ($\simeq$~10~MeV/c)
and with a small gap between the intervals ($\simeq$~10~MeV/c). 

A polarization growth with the final state proton momentum at $K>$~1530 MeV/c was
found. A structure in the polarization and cross section data in this region
was observed for the first time. The structure is possibly related to
a proton quasielastic scattering off the two-nucleon, three-nucleon, and 
four-nucleon correlations.
    
   The authors are grateful to the PNPI 1~GeV proton accelerator staff
for the stable beam operation.
Also, the authors would like to express their gratitude to A.A.~Vorobyov and S.L.~Belostotski
for their support and fruitful discussions. We thank D.A.~Prokofiev for his valuable
help in this paper preparation.\\
\\
\\
\begin{table}
\caption{The
polarization P of the scattered proton in the reaction
$^{40}$Ca({\it p, p'})X  at 1~GeV and lab. angle
$\Theta$=21$^\circ$} 
\label{table:pol-ca}
\begin{tabular}{c|c||c|c||c|c}
\hline
$K$ & $P$ & $K$ & $P$ & $K$ & $P$  \\
MeV/c &  & MeV/c &  & MeV/c &   \\
\hline
1417.3 & 0.217$\pm$0.018 & 1502.8 & 0.274$\pm$0.011 & 1592.9 & 0.364$\pm$0.015
\\
1427.0 & 0.272$\pm$0.018 & 1508.8 & 0.303$\pm$0.019 & 1605.1 & 0.387$\pm$0.022
\\
1436.7 & 0.217$\pm$0.017 & 1519.1 & 0.281$\pm$0.018 & 1618.4 & 0.361$\pm$0.014
\\
1446.6 & 0.233$\pm$0.017 & 1529.4 & 0.298$\pm$0.019 & 1629.6 & 0.361$\pm$0.015
\\
1455.5 & 0.234$\pm$0.011 & 1538.9 & 0.306$\pm$0.013 & 1641.6 & 0.368$\pm$0.019
\\
1465.4 & 0.275$\pm$0.012 & 1549.3 & 0.302$\pm$0.013 & 1652.7 & 0.397$\pm$0.022
\\
1474.8 & 0.296$\pm$0.014 & 1559.1 & 0.322$\pm$0.017 & 1664.0 & 0.430$\pm$0.026
\\
1484.8 & 0.291$\pm$0.014 & 1571.3 & 0.310$\pm$0.013 &        &
\\
1495.0 & 0.304$\pm$0.015 & 1582.0 & 0.308$\pm$0.013 &        &
\\
\hline
\end{tabular}
\end{table}       
 \\
 \\
 \\                                      
 \begin{table}
\caption{The
polarization P of the scattered proton in the reaction
$^{12}$C({\it p, p'})X  at 1~GeV and lab. angle
$\Theta$=21$^\circ$}
\label{table:pol-c}
\begin{tabular}{c|c||c|c||c|c}
\hline
$K$ & $P$ & $K$ & $P$ & $K$ & $P$  \\
MeV/c &  & MeV/c &  & MeV/c &   \\
\hline
1407.9 & 0.216$\pm$0.027 & 1483.8 & 0.297$\pm$0.030 & 1571.6 & 0.301$\pm$0.015
\\
1417.4 & 0.295$\pm$0.026 & 1496.9 & 0.279$\pm$0.018 & 1582.4 & 0.294$\pm$0.017
\\
1427.0 & 0.313$\pm$0.025 & 1507.3 & 0.278$\pm$0.017 & 1593.3 & 0.296$\pm$0.018
\\
1436.7 & 0.257$\pm$0.024 & 1518.8 & 0.268$\pm$0.021 & 1605.1 & 0.369$\pm$0.025
\\
1446.5 & 0.224$\pm$0.024 & 1529.1 & 0.297$\pm$0.022 & 1618.5 & 0.351$\pm$0.016
\\
1455.5 & 0.243$\pm$0.018 & 1538.9 & 0.259$\pm$0.015 & 1629.7 & 0.337$\pm$0.016
\\
1465.3 & 0.279$\pm$0.020 & 1549.3 & 0.282$\pm$0.017 & 1641.6 & 0.368$\pm$0.022
\\
1473.8 & 0.285$\pm$0.030 & 1559.3 & 0.267$\pm$0.022 & 1652.9 & 0.384$\pm$0.024
\\
\hline
\end{tabular}
\end{table}
\clearpage
\begin{table}
\caption{The
cross section of the reaction $^{40}$Ca({\it p, p'})X at 1~GeV
and lab. angle $\Theta$=21$^\circ$ } 
\label{table:cross-ca}

\begin{tabular}{c|c||c|c||c|c}
\hline
$K$ & $\frac{d^2\sigma}{d\Omega dK}$ & $K$ & $\frac{d^2\sigma}{d\Omega dK}$ & $K$ & $\frac{d^2\sigma}{d\Omega dK}$  \\
MeV/c & mb/(sr$\cdot$MeV/c)  & MeV/c & mb/(sr$\cdot$MeV/c) & MeV/c & mb/(sr$\cdot$MeV/c)  \\
\hline
1375.0 & .3085$\pm$.0032 & 1475.1 & .4928$\pm$.0047 & 1574.9 & .2814$\pm$.0036
\\
1385.0 & .3283$\pm$.0034 & 1485.0 & .4922$\pm$.0049 & 1584.9 & .2659$\pm$.0070
\\
1395.1 & .3547$\pm$.0034 & 1495.0 & .4918$\pm$.0047 & 1594.9 & .2271$\pm$.0060
\\
1405.1 & .3659$\pm$.0038 & 1505.0 & .4956$\pm$.0029 & 1604.9 & .1950$\pm$.0028
\\
1415.0 & .3921$\pm$.0107 & 1515.0 & .4760$\pm$.0040 & 1614.9 & .1660$\pm$.0028
\\
1425.0 & .4083$\pm$.0051 & 1525.0 & .4560$\pm$.0043 & 1625.0 & .1585$\pm$.0082
\\
1435.1 & .4294$\pm$.0053 & 1535.0 & .4347$\pm$.0043 & 1634.9 & .1331$\pm$.0015
\\
1445.1 & .4385$\pm$.0055 & 1545.0 & .4023$\pm$.0025 & 1644.9 & .1090$\pm$.0015
\\
1455.1 & .4750$\pm$.0034 & 1554.9 & .3793$\pm$.0076 & 1654.9 & .0788$\pm$.0021
\\
1465.1 & .4932$\pm$.0043 & 1565.0 & .3197$\pm$.0036 & 1664.9 & .0622$\pm$.0019
\\
\hline
\end{tabular}

\end{table}
\begin{table}
\caption{The
cross section of the reaction $^{12}$C({\it p, p'})X at 1~GeV
and lab. angle $\Theta$=21$^\circ$}
\label{table:cross-c}
\begin{center}
\begin{tabular}{c|c||c|c||c|c}
\hline
$K$ & $\frac{d^2\sigma}{d\Omega dK}$ & $K$ & $\frac{d^2\sigma}{d\Omega dK}$ & $K$ & $\frac{d^2\sigma}{d\Omega dK}$ 
\\
MeV/c & mb/(sr$\cdot$MeV/c)  & MeV/c & mb/(sr$\cdot$MeV/c) & MeV/c & mb/(sr$\cdot$MeV/c) \\
\hline
1375.1 & .1774$\pm$.0018 & 1475.0 & .2677$\pm$.0024 & 1574.9 & .1272$\pm$.0016
\\
1385.1 & .1911$\pm$.0020 & 1485.0 & .2676$\pm$.0025 & 1584.9 & .1186$\pm$.0029
\\
1395.1 & .2045$\pm$.0021 & 1495.0 & .2676$\pm$.0024 & 1594.9 & .1016$\pm$.0013
\\
1405.0 & .2197$\pm$.0023 & 1505.0 & .2723$\pm$.0023 & 1604.9 & .0829$\pm$.0017
\\
1415.0 & .2357$\pm$.0056 & 1515.0 & .2573$\pm$.0018 & 1614.9 & .0722$\pm$.0017
\\
1425.0 & .2373$\pm$.0025 & 1525.0 & .2410$\pm$.0019 & 1624.9 & .0703$\pm$.0041
\\
1435.0 & .2494$\pm$.0026 & 1535.0 & .2159$\pm$.0017 & 1635.0 & .0618$\pm$.0009
\\
1445.0 & .2562$\pm$.0028 & 1544.9 & .2061$\pm$.0011 & 1644.9 & .0476$\pm$.0011
\\
1455.0 & .2692$\pm$.0017 & 1554.9 & .1810$\pm$.0037 & 1655.0 & .0349$\pm$.0012
\\
1465.0 & .2677$\pm$.0023 & 1564.9 & .1533$\pm$.0015 &        &
\\
\hline
\end{tabular}
\end{center}
\end{table}

\end{document}